\documentclass{aastex63}

\received{November 20, 2019}
\revised{July 28, 2020}
\accepted{August 7, 2020}
\submitjournal{ApJ}

\shorttitle{Foreshock transients and secondary shocks}
\shortauthors{An et al.}

\usepackage{amsmath}
\usepackage{amssymb}
\usepackage{graphicx}
\usepackage{hyperref}
\usepackage{xcolor}

\begin{document}

\title{Formation of foreshock transients and associated secondary shocks}

\correspondingauthor{Xin An, Terry Z.~Liu}
\email{xinan@atmos.ucla.edu, terryliuzixu@ucla.edu}

\author{Xin An}
\affiliation{Department of Atmospheric and Oceanic Sciences, University of California, Los Angeles, California 90095, USA.}

\author{Terry Z.~Liu}
\affiliation{Department of Earth, Planetary and Space Sciences, University of California, Los Angeles, California 90095, USA.}
\affiliation{University Corporation for Atmospheric Research, Boulder, Colorado 80307, USA.}

\author{Jacob Bortnik}
\affiliation{Department of Atmospheric and Oceanic Sciences, University of California, Los Angeles, California 90095, USA.}

\author{Adnane Osmane}
\affiliation{Department of Physics, University of Helsinki, Helsinki 00014, Finland.}

\author{Vassilis Angelopoulos}
\affiliation{Department of Earth, Planetary and Space Sciences, University of California, Los Angeles, California 90095, USA.}

\begin{abstract}

Upstream of shocks, the foreshock is filled with hot ions. When these ions are concentrated and thermalized around a discontinuity, a diamagnetic cavity bounded by compressional boundaries, referred to as a foreshock transient, forms. Sometimes, the upstream compressional boundary can further steepen into a secondary shock, which has been observed to accelerate particles and contribute to the primary shock acceleration. However, secondary shock formation conditions and processes are not fully understood. Using particle-in-cell simulations, we reveal how secondary shocks are formed. From 1D simulations, we show that electric fields play a critical role in shaping the shock's magnetic field structure, as well as in coupling the energy of hot ions to that of the shock. We demonstrate that larger thermal speed and concentration ratio of hot ions favors the formation of a secondary shock. From a more realistic 2D simulation, we examine how a discontinuity interacts with foreshock ions leading to the formation of a foreshock transient and a secondary shock. Our results imply that secondary shocks are more likely to occur at primary shocks with higher Mach number. With the secondary shock's previously proven ability to accelerate particles in cooperation with a planetary bow shock, it is even more appealing to consider them in particle acceleration of high Mach number astrophysical shocks.

\end{abstract}

\keywords{Plasma astrophysics, Shocks, Planetary bow shocks}


\section{Introduction}\label{sec-introduction}
Although shocks are among the most powerful particle accelerators in the universe, how they accelerate particles to extreme energies, e.g., generate cosmic rays, is still not fully understood \citep[see review in][]{treumann2009fundamentals}. Recent in-situ observations at the Earth's bow shock have shown that nonlinear structures in the foreshock can play an important role in shock acceleration \citep{liu2016observations, wilson2016relativistic, liu2017statistical, liu2017fermi, liu2018ion, turner2018autogenous, liu2019relativistic}. A region upstream of the shock, the foreshock, is magnetically connected to the shock and filled with particles coming from it \citep{eastwood2005foreshock}. The foreshock is very dynamic and there are many nonlinear transient structures with large plasma and field fluctuations referred to as foreshock transients. Because of the density and magnetic field variations and/or plasma deflection, foreshock transients can result in dynamic pressure perturbations. When foreshock transients convect towards and connect with the planetary bow shock, the bow shock surface can be disturbed. Such perturbation can propagate to the magnetosheath, the magnetopause and thus the magnetosphere \citep[e.g.,][]{sibeck1999comprehensive, turner2011multispacecraft}. Hot flow anomalies (HFAs) \citep{schwartz1985active, thomsen1986hot, thomsen1988origin, omidi2007formation} and foreshock bubbles (FBs) \citep{omidi2010foreshock, turner2013first} are two most significant types of foreshock transients due to their large spatial scale (from several foreshock ion gyroradii to almost the entire width of the ion foreshock) and the significant plasma perturbations they entail. Foreshock transients have a lifetime of several minutes. Their main characteristic is a hot, diamagnetic cavity surrounded by a compressional boundary or a secondary shock [Figs.~\ref{setup}(a) and \ref{setup}(b)]. 

Recent observations found that foreshock transients, especially HFAs and FBs, can accelerate particles and contribute to the primary shock acceleration \citep[e.g.,][]{wilson2013shocklets, liu2017statistical}. As foreshock transients convect with the upstream flow, particles enclosed within their boundary and the primary shock can experience Fermi acceleration \citep{liu2017fermi, liu2018ion, turner2018autogenous}. Secondary shocks have also been observed to accelerate upstream particles on their own through the shock drift acceleration and even to form a secondary foreshock \citep{liu2016observations}. Secondary shocks can also capture and further energize primary shock-accelerated electrons through betatron acceleration \citep{liu2019relativistic}. Recent observations also identified magnetic reconnection inside foreshock transients contributing to the electron acceleration/heating \citep{liu2020magnetospheric}. These additional accelerations by foreshock transients and their secondary shocks provide a possible solution to Fermi's `injection problem' (an unresolved seed population of energetic particles for further acceleration \citep{treumann2009fundamentals}) and increase the acceleration efficiency of primary shocks. To date foreshock transients have only been observed at planetary bow shocks where in-situ observations are available \citep[e.g.,][]{masters2009hot, turner2013first, collinson2015hot}. Whether foreshock transients exist at high Mach number astrophysical shocks, such as supernova-driven shocks, is unknown or can only be inferred from simulations \cite[e.g.,][]{giacalone2010interaction}.  Therefore, to infer their existence and contribution at other shock systems, it is important to fully understand and quantify the formation process of foreshock transients.

Based on various simulations, conceptual models have been proposed for the formation of HFAs and FBs. Test particle simulations by \citet{burgess1989effect} show that when convection electric field points towards the tangential discontinuity (TD) that intersects a shock, specularly reflected ions either traveling away upstream (quasi-parallel shock) or not (quasi-perpendicular shock) can be trapped by the TD and channeling along it. Hybrid simulations \citep[e.g.,][]{thomas1991hybrid, lin2002global, omidi2007formation} show that such scenario can lead to the formation of HFAs. In order to have sufficient time to trap foreshock ions and form an HFA, TDs need to move along the bow shock surface very slowly as shown in the statistical study by \citet{schwartz2000conditions}. Additionally, hybrid simulations by \citet{omidi2010foreshock} show that when field-aligned foreshock ions cross a rotational discontinuity (RD), an FB can form. To understand how the interaction between foreshock ions and a TD/RD leads to the formation of an HFA or FB, \citet{archer2015global} and \citet{liu2015themis} suggest that, because of the magnetic field direction change across the discontinuity, part of the foreshock ion's parallel speed projects onto the perpendicular direction increasing the thermal energy in the perpendicular direction (thermalization). The decrease of the parallel speed also results in the increase of the foreshock ion density (concentration), due to conservation of mass flux. The thermalization and concentration of foreshock ions increase their thermal pressure resulting in an expansion and formation of a foreshock bubble or a hot flow anomaly. However, such a model is too qualitative. For example, because foreshock ion gyradii are comparable to the system size, the concept of thermal pressure of foreshock ions is invalid. The role of electrons in the formation process is neglected. Under what condition there will be a secondary shock is unknown. 

The rest of the paper is organized as follows. In Section \ref{sec-1d-pic}, we will study a simple initial value problem using particle-in-cell (PIC) simulations in one spatial dimension. In this configuration, we initialize a layer of hot ions and simulate the plasma expansion perpendicular to the magnetic field and the resulting formation of foreshock transients. We will investigate the detailed roles of hot foreshock ions, cold ambient ions, electrons, and the associated electromagnetic field during the formation process. We will further reveal the critical parameters that determine the secondary shock formation. In Section \ref{sec-2d-pic}, we will study a more realistic boundary value problem using a PIC simulation in two spatial dimensions. In this configuration, we continuously inject energetic ions at a boundary and simulate the interaction of injected ions with a rotational discontinuity and the formation of a foreshock transient and a secondary shock. We will compare the 2D simulation with 1D simulations and show effects of the additional spatial dimension. In Section \ref{sec-conclusions-discussions}, we will summarize and discuss our results.

\section{Expansion perpendicular to the magnetic field: 1D PIC simulations}\label{sec-1d-pic}
\begin{figure*}[tphb]
 \centering
 \includegraphics[width=6.75in]{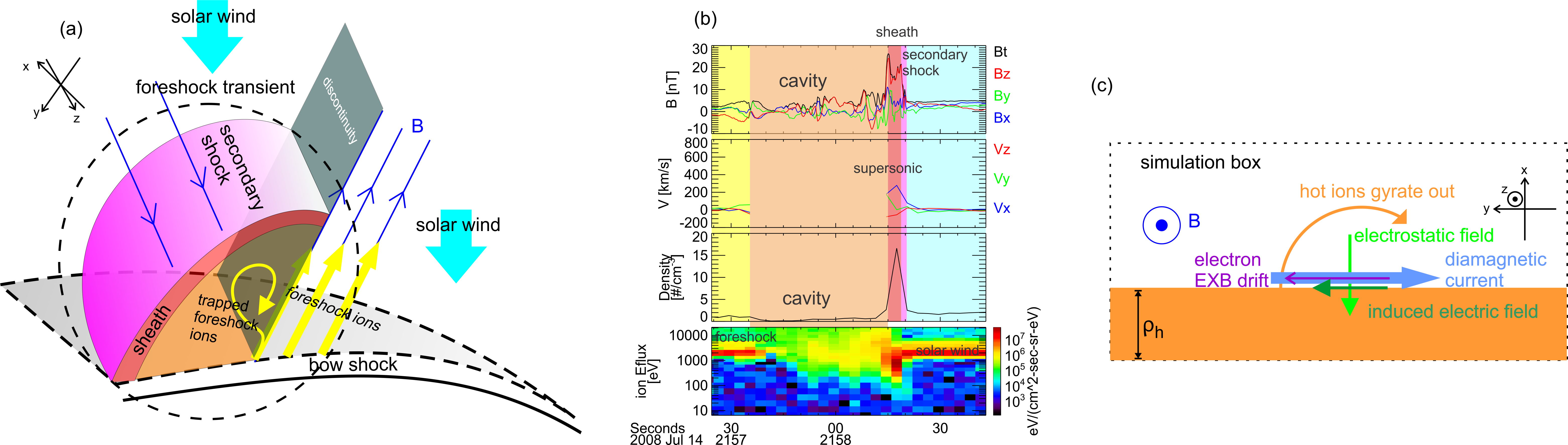}%
 \caption{\label{setup}A foreshock transient and its associated secondary shock. (a) A sketch of the foreshock transient upstream of the bow shock. On the right side of the discontinuity (gray board), foreshock ions (yellow arrows) move along magnetic field lines (blue arrows). Because of their large gyroradius, some foreshock ions gyrate across the discontinuity instead (curved yellow arrow), are trapped and become concentrated and thermalized (orange region), resulting in a fast expansion that forms a secondary shock (purple surface). (b) A typical observation of a foreshock transient upstream of the Earth's bow shock \citep{turner2013first, wilson2016relativistic, liu2019relativistic}. From top to bottom are magnetic field, ion bulk velocity in the solar wind rest frame, plasma density, and ion energy flux spectrum in the Earth's rest frame. Here the $x$ direction is defined as the normal direction of the secondary shock; the $z$ direction is defined as the direction of maximum variation of the magnetic field; and the $y$ direction completes the right-handed coordinate system (as sketched in (a)). The colors of the shaded regions correspond to those shown in (a). (c) A sketch of the simulation setup. The coordinate system represents the geometry in (a) and (b).}
 \end{figure*}

To explore the foreshock transient formation, we carried out a series of PIC simulations using the \textsc{osiris} code \citep{fonseca2002osiris, hemker2015particle}, including $1$ spatial ($x$) and $3$ velocity components ($v_x$, $v_y$, $v_z$). All the runs were in the rest frame of the upstream flow, so there is no background flow in the initial setup (see Appendix \ref{appendix-computational-setup} for details of the simulation setup). A uniform background magnetic field $B_0$ is oriented in the $+z$ direction [Fig.~\ref{setup}(c)]. The initial plasma density $n_0$ is uniform in the computation domain. In the foreshock region, when ions backstreaming from the primary shock encounter certain discontinuities in the upstream flow, they are concentrated and thermalized \citep{archer2015global, liu2015themis}. To mimic the concentration of hot ions, a mixture of hot and ambient ions (and electrons) is initialized in layer $0 \leqslant x \leqslant \rho_h$ [Fig.~\ref{setup}(c)], where $\rho_{h}$ is the gyroradius corresponding to the initial thermal velocity $v_{Th}$ of hot ions. Inside this layer, the densities of hot ions, ambient ions and electrons are $0.2 n_0$, $0.8 n_0$, and $n_0$, respectively; outside it, there are only ambient ions and electrons, each with density $n_0$. The hot ions have an initial thermal velocity of $v_{Th} = 9 v_A$; the ambient ions and electrons have initial thermal velocities of $v_{Ti} = 0.3 v_A$ and  $v_{Te} = 3 v_A$, respectively. Here the Alfv\'en velocity is $v_A = B_0 / \sqrt{4 \pi n_0 m_i}$, and the speed of light is $c = 300 v_A$. The highly nonequilibrium state of the initial setup is consistent with the observed characteristics of different particle species at an early stage (less than an ion cyclotron period from the birth of the hot ion core) of foreshock transient formation \citep{turner2013first, omidi2010foreshock}. Given the reduced ion-to-electron mass ratio $m_i / m_e = 100$, the ordering of the typical gyroradii of the three species is $\rho_{h} = 30 \rho_i = 300 \rho_e$. The wide gap between the gyroradii of the hot ions and those of the ambient ions and electrons, which captures the essential ordering in spacecraft observations, will have a significant effect on the evolution of the system. Later we also discuss runs that vary the initial thermal velocity and the concentration ratio of hot ions.

\subsection{Formation process}\label{sec-formation-process}
\begin{figure}[tphb]
 \centering
 \includegraphics[width=6.75in]{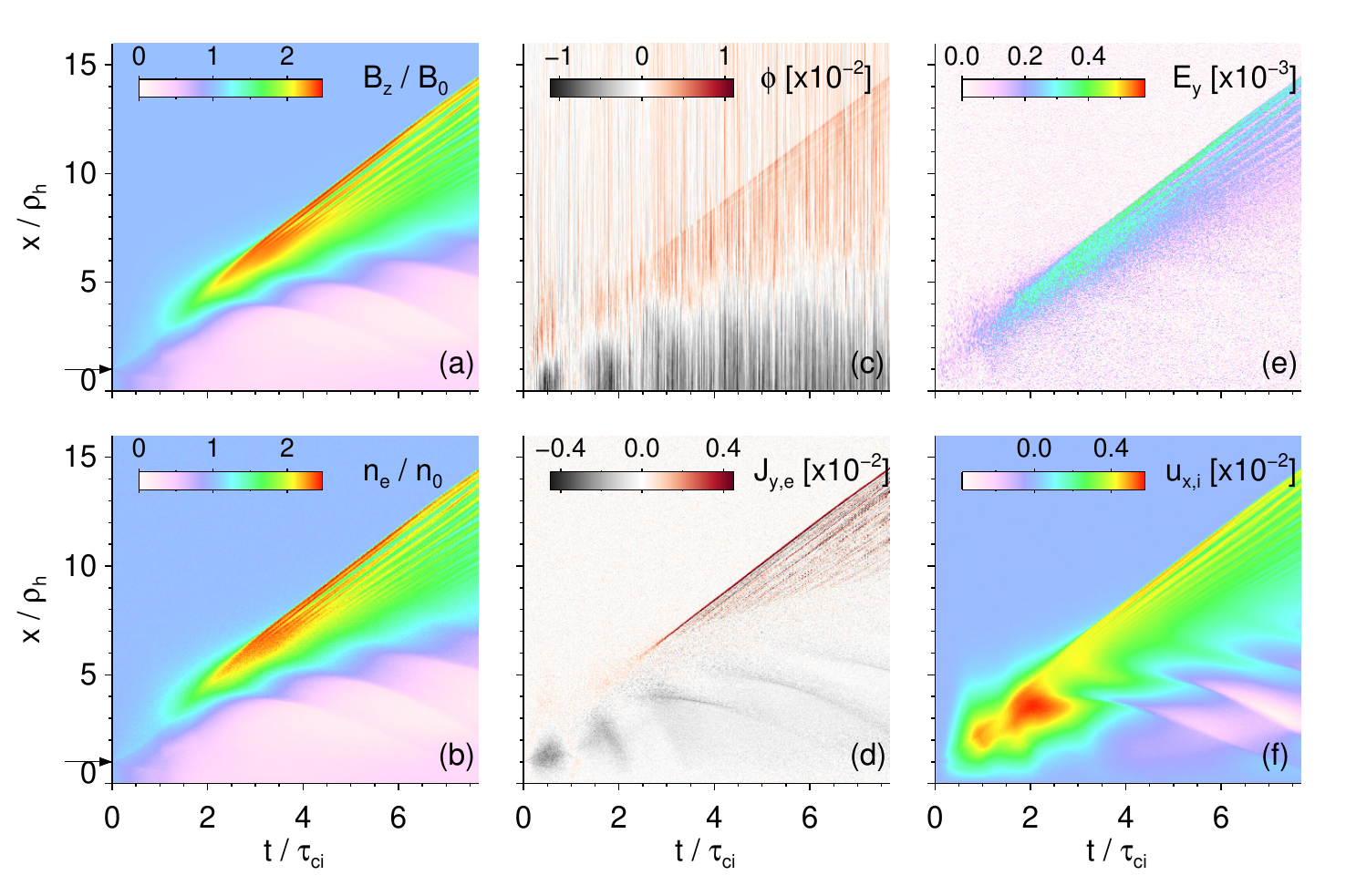}%
 \caption{\label{fld}The formation of a foreshock transient and a secondary shock. (a)-(f) The spatio-temporal evolution of relevant field and particle quantities. The position $x$ is normalized to $\rho_h$. The time $t$ is normalized to $\tau_{ci}$. The arrows next to (a) and (b) indicate the original boundary between hot and ambient plasmas. (a) The magnetic field $B_z$ in the $z$ direction. (b) The electron plasma density $n_e$. The evolution of $n_e$ is identical to that of $B_z$ because their transport equations have the same form in 1D. (c) The electrostatic potential $\phi$. The corresponding electric field, $E_x = -\partial \phi / \partial x$, resides in the large electron density gradient between the magnetic cavity and the compressional boundary. (d) The electron current $J_{y,e}$ in the $y$ direction. (e) The induction electric field $E_y$. (f) The fluid velocity of the ambient ions $u_{x,i}$ in the $x$ direction. In panels (c)-(f) and hereafter, the electric potential has the dimension $m_e c^2 / e$, the current density has the dimension $n_0 e c$, the electric field has the dimension $\frac{m_e c^2}{e \cdot c / \omega_{pe}}$, and the velocity has the dimension $c$.}
 \end{figure}

Here we explain the details of the foreshock transient formation process. In our simulations, the concentrated hot ions drive the system to form a diamagnetic cavity and a secondary shock [Fig.~\ref{fld}]. A fundamental question is, how is the energy of the hot ions transferred to the magnetic field and ambient plasma flows tied to understanding of electric fields, which control energy transfer between particles and fields. And, indeed, two types of electric fields were identified in the simulations: electrostatic fields in the $-x$ direction [Fig.~\ref{fld}(c)] and induction electric fields in the $+y$ direction [Fig.~\ref{fld}(e)]. As the energetic ions move out of the hot layer across the magnetic field in the first gyration ($0 < t < \tau_{ci}$, $\tau_{ci}$ being the ion cyclotron period in $B_0$), an electrostatic field $E_x$ directed from these displaced hot ions to the initial hot layer (i.e., $E_x < 0$) is generated. Correspondingly, the electrostatic potential decreases in the hot layer and is enhanced outside it [Fig.~\ref{fld}(c)]. This electrostatic field causes a drift $u_{y,e} = -c E_x / B_z$ of electrons, an electron current in the $-y$ direction [Figs.~\ref{fld}(d) and \ref{setup}(c)]. The electron current and the hot ion (partial diamagnetic) current (due to the density gradient and partial gyration of hot ions at the interface), both flowing in the $-y$ direction, cannot be offset by the ambient ion (partial diamagnetic) current (flowing in the opposite, $+y$, direction). The resulting net current, Hall current, (see Appendix \ref{appendix-diamagnetic-current} for the ion diamagnetic current and total current), reduces the magnetic field on one side (the cavity) and enhances it on the other side (the compressional boundary) [Fig.~\ref{fld}(a)]. Simultaneously, an induction electric field $E_y$ is generated by these magnetic field variations [Figs.~\ref{fld}(e) and \ref{setup}(c)], in order to keep the electrons drifting across the magnetic field region extending in the $+x$ direction, encompassing the hot ions that gyrate out. Ambient ions, whose gyration is completely immersed in the interaction region of $2 \rho_h$ due to their smaller gyroradius, also gain momentum in the $+x$ direction via the induction electric field, eventually forming a streaming flow with velocity $u_{x, i} = c E_y / B_z$ (see Fig.~\ref{fld}(f) and Supplemental Video 1\footnote{Version 1.0 of Supplemental Video 1 is archived  on Zenodo \url{https://doi.org/10.5281/zenodo.3951168}.}). Their acceleration is similar to the ion pick-up process at a comet \citep[e.g.,][]{biermann1967interaction, gloeckler1986cometary}. As a result, the ambient plasma, as well as the magnetic flux, are transported outward, and thus the density and magnetic field strength are depleted on one side, forming a cavity and piled up on the other side, forming the compressional boundary (see Figs.~\ref{fld}(a), \ref{fld}(b) and Supplemental Video 2\footnote{Version 1.0 of Supplemental Video 2 is archived  on Zenodo \url{https://doi.org/10.5281/zenodo.3951168}.}). This cycle repeats until the compressional boundary finally detaches from the hot ions at $3\tau_{ci} \lesssim t < 4\tau_{ci}$ (see Supplemental Video 3\footnote{Version 1.0 of Supplemental Video 3 is archived  on Zenodo \url{https://doi.org/10.5281/zenodo.3951168}.}). As explained below, the energy exchange between the fields and the hot ions ceases at the time of this detachment. The density of hot ions decreases in surrounding space, but remains concentrated in the cavity region. The compressional boundary, which continues to move at supermagnetosonic speed, steepens into a shock with the Mach number $M_f = 2.2$ ($M_f$, defined as the ratio of the upstream flow speed in the shock normal incidence frame to the local fast magnetosonic speed $v_f$).

\begin{figure}[tphb]
 \centering
 \includegraphics{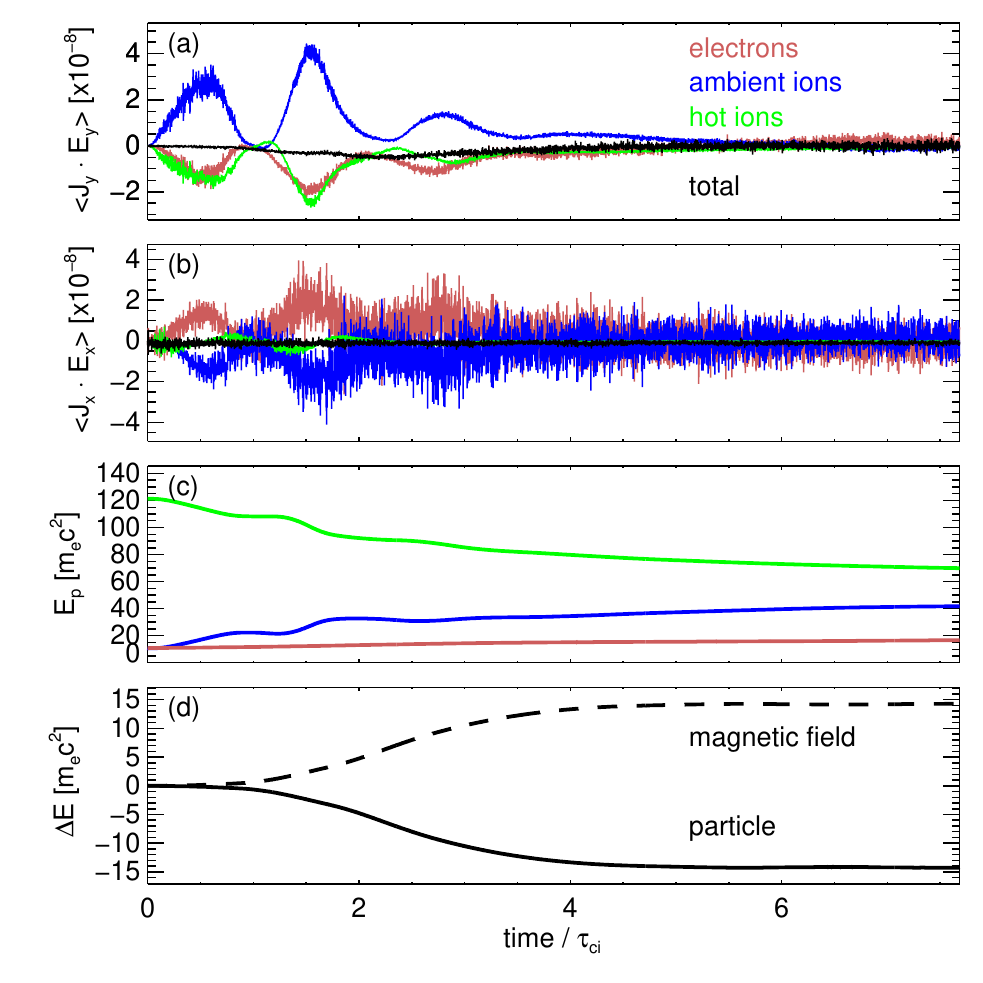}%
 \caption{\label{ene}Energy exchange between fields and particles. (a),(b) The history of the rate of work done by $E_x$ and $E_y$ on electrons (red), ambient ions (blue), hot ions (green), and all particles (black). The spatial average of any quantity $Q$ is defined by $\langle Q \rangle = \frac{1}{L_x}\int_0^{L_x} Q(x) \,dx$, where $L_x$ is the domain size. (c) The kinetic energy of electrons (red), ambient ions (blue), and hot ions (green) with respect to time. (d) The net change in total kinetic energy (solid) and magnetic field energy (dashed) with respect to time. Compared with the magnetic field energy, the electric field energy is negligible.}
 \end{figure}

Details of the energy transfer between fields and particles are shown in Fig.~\ref{ene}. Each time the hot ions gyrate out to the ambient plasma in the time interval $0 < t < 3 \tau_{ci}$, a surge in the rate of work done by the electric field $E_y$ and $E_x$ on the particles is seen [Figs.~\ref{ene}(a) and \ref{ene}(b)], which is associated with an energy decrease in the hot ions and an energy increase in the ambient ions and electrons [Fig.~\ref{ene}(c)]. In this process, the hot ions transfer energy to the fields dominantly through the partial gyration against the induction electric field $E_y$ ($J_{y, h} \cdot E_y < 0$), whereas the work done by the electrostatic field $E_x$ on hot ions is relatively small, because the spatial location of $E_x$ lags behind that of the hot ion current $J_{x,h}$. The electron current $J_{y,e}$ is a generator ($J_{y,e} \cdot E_y < 0$), through which electrons also transfer energy to magnetic fields. The electrostatic field $E_x$ tends to accelerate electrons [Fig.~\ref{ene}(b)], however, leading to a net electron energy increase [Fig.~\ref{ene}(c)], which is consistent with a recent statistical study showing that electrons are almost always heated/accelerated inside foreshock transients \citep{liu2017statistical}. Accelerated by the induction electric field $E_y$ [Fig.~\ref{ene}(a)], ambient ions gain energy in the form of streaming in the $+x$ direction [Fig.~\ref{fld}(f)]. In this way, the induction electric field mediates momentum and energy coupling between hot and ambient ions. We also note that in addition, the electrostatic field $E_x$ creates the electron current and thus plays a critical role in shaping the magnetic cavity and compression. These electric fields have also been observed in other scenarios, such as solar wind-barium interaction in the Active Magnetospheric Particle Tracer Explorers (AMPTE) experiment \citep[e.g.,][]{papadopoulos1987collisionless} and laser-produced plasma expansion in laboratory experiments \citep[e.g.,][]{bonde2015electrostatic, bondarenko2017collisionless}. Because the decrease in the total kinetic energy of particles is matched by the increase in the magnetic field energy [Fig.~\ref{ene}(d)], global energy is conserved. The energy transfer between fields and particles almost vanishes after $t = 4 \tau_{ci}$, as hot ions are detached from the outward-propagating compressional boundary (i.e., the region where $E_y$ is located).

\begin{figure}[tphb]
	\centering
	\includegraphics[width=5in]{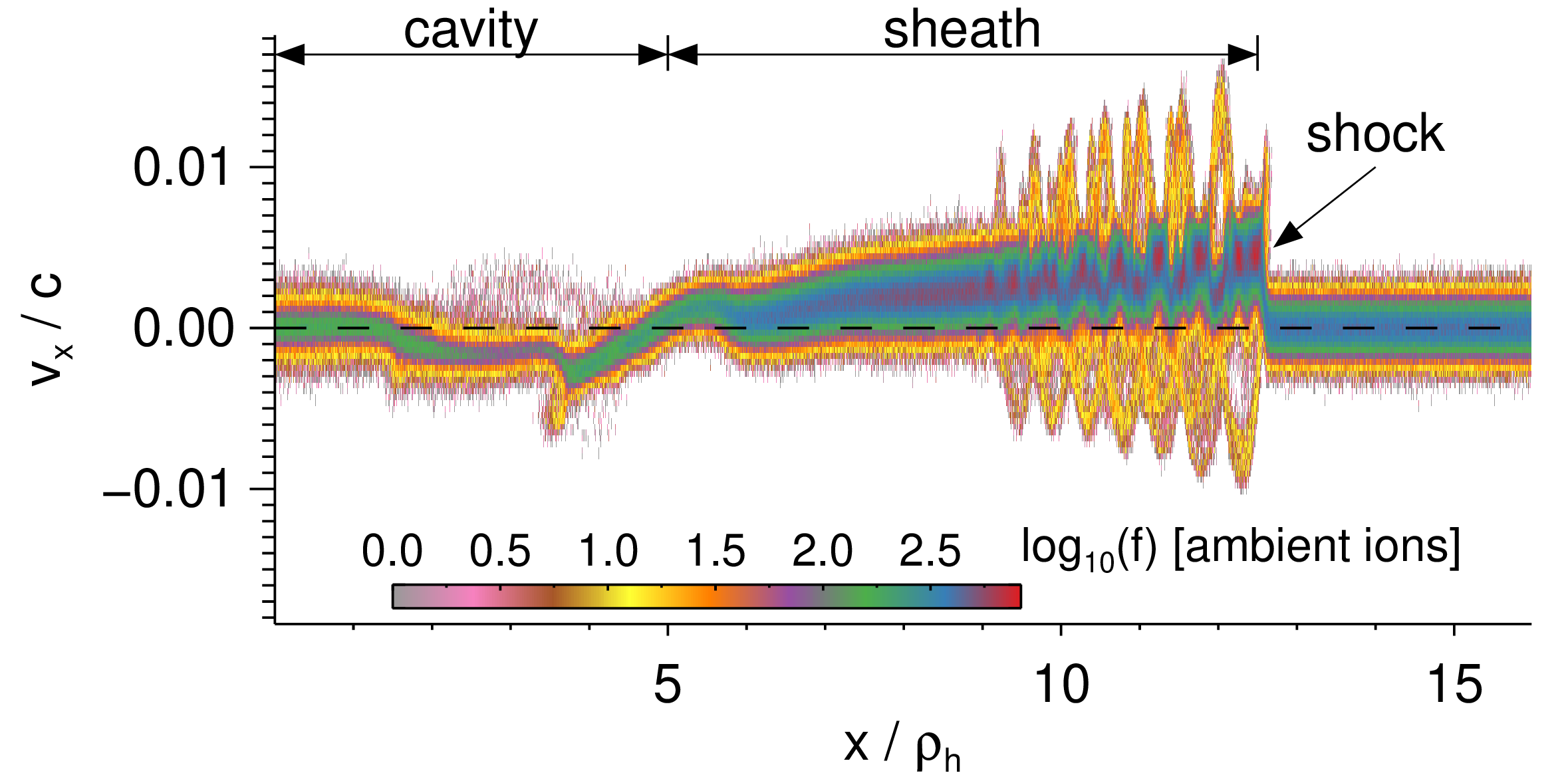}%
	\caption{\label{oscillations}The phase space portrait of ambient ions at $t = 6.5 \tau_{ci}$. Colors denote the phase space density of ambient ions. At this instant, the electromagnetic oscillations are located at $9 < x / \rho_h < 13$, and the return flow from the sheath to the cavity is located at $2 < x / \rho_h < 5$. The thin dashed line at $v_x = 0$ indicates which part of the phase space is flowing towards/away from the center.}
\end{figure}

When swept over by the shock, ambient ions from the upstream are accelerated by the finite $E_y$ and begin to gyrate in large radii. These transmitted ions have gyrating velocities comparable to the sheath flow velocity, as shown in Figure \ref{oscillations}. The transmitted ions are non-gyrotropic \citep[e.g.,][]{sckopke1983evolution, sckopke1990ion, burgess1989ion} and therefore causes magnetic oscillations along the background magnetic field [Fig.~\ref{fld}(a)], which is consistent with theoretical predictions \citep{gedalin2015collisionless} and spacecraft observations \citep{pope2019first}. In fact, electric oscillations of both $E_x$ (inferred from the oscillations of electron current shown in Fig.~\ref{fld}(d)) and $E_y$ [Fig.~\ref{fld}(e)] are also present due to the non-gyrotropy of transmitted ions. Due to gyrophase mixing, the electromagnetic oscillations gradually decrease in amplitude further downstream from the shock ramp. The electromagnetic oscillations significantly disturb the ion flow and heat the ambient ions in the sheath region (see Fig.~\ref{oscillations}), leading to dissipation of the shock structure. In the meantime, magnetic flux is slowly transported from the sheath to the cavity, as seen in the return flow ($u_{x, i} < 0$) of ambient ions [Figs.~\ref{fld}(f) and \ref{oscillations}] and electrons (not shown). Eventually the magnetic field in the cavity is expected to approach its ambient value $B_0$, i.e., the cavity will vanish.

\subsection{Parameter scans}\label{sec-parameter-scans}
To determine the conditions necessary for secondary shock formation and to connect it to the Mach number of the primary shock, we perform a parameter scan of the thermal velocity and the concentration ratio of hot ions [Fig.~\ref{scan}]. Looking into the evolution of the system in each run (see Appendix \ref{appendix-parameter-scan} for the magnetic field structure in each run), we see that the large thermal velocities and high concentration ratios of hot ions favor secondary shock formation. In the simulations, at the lower limit of hot ion thermal velocity and concentration, weak magnetic bumps, rather than secondary shocks, appear and propagate approximately at the fast magnetosonic speed of the ambient plasma. As either the thermal velocity or the concentration of hot ions is increased, shock structures emerge, and both the Mach number and the magnetic compression ratio of these shocks increase [Figs.~\ref{scan}(a) and \ref{scan}(b)]. The relation between the Mach number and the magnetic compression ratio for the simulated secondary shocks agrees with that derived from Rankine-Hugoniot relation [Fig.~\ref{scan}(c)]. On the one hand, because the hot ion/foreshock ion speed is proportional to the upstream flow speed, the ratio of hot ion thermal speed to the fast magnetosonic speed is proportional to the Mach number of the primary shock \citep{burgess2012ion}. The ratio of the foreshock ion density to the incident upstream ion density, on the other hand, also increases with the Mach number of the primary shock up to $\sim 0.2$ as the Mach number of the primary shock exceeds $\sim 6$ \citep{paschmann1983ion}. Because both the thermal velocity and the concentration ratio of hot ions are positively correlated with the Mach number of the primary shock, the parameter scan strongly indicates that the secondary shocks are more likely formed at high Mach number shocks. This result is consistent with the statistical study \citep{liu2017statistical}, which shows that foreshock transients are more likely to occur when the Mach number of the Earth's bow shock is higher (see Appendix \ref{appendix-correlation} for the statistical results). Multiple case study by \citet{turner2020microscopic} also shows that the expansion speed of FBs is positively correlated with the primary shock Mach number.

 \begin{figure}[tphb]
 \centering
 \includegraphics{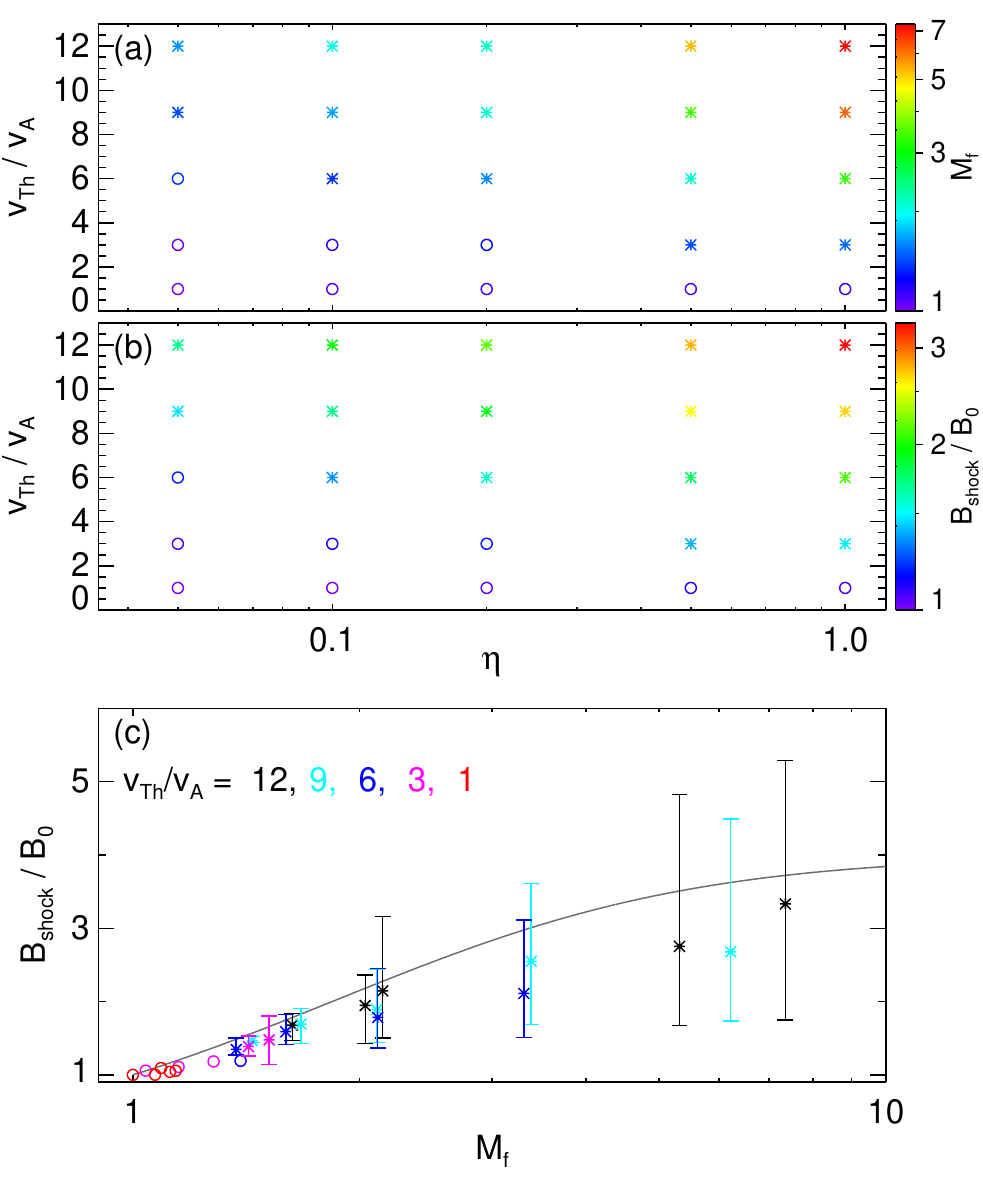}%
 \caption{\label{scan}Parameter scan examining the conditions under which secondary shocks form. (a) The Mach number of secondary shocks and (b) the magnetic compression ratio across the secondary shock, as a function of the hot ion thermal velocity $v_{Th} = (12 v_A, 9 v_A, 6 v_A, 3 v_A, v_A)$ for a given concentration ratio of hot ions $\eta = (0.05, 0.1, 0.2, 0.5, 1.0)$. The Mach number and magnetic compression ratio are color coded for each data point. The asterisks indicate the runs in which a supermagnetosonic shock is formed, whereas the circles indicate the runs in which no apprimary shock is formed. (c) The magnetic compression ratio with respect to the Mach number of secondary shocks in the parameter scan. In evaluating the magnetic compression ratio for each run, we use the magnetic field averaged over the compressional boundary. The upper and lower bound of the error bars denote the maximum and minimum magnetic fields in the compressional boundary, respectively. The solid line represents the predicted jump condition of perpendicular shocks based on magnetohydrodynamic conservation laws.}
 \end{figure}

\section{Expansion in two spatial dimensions: Injection at a discontinuity}\label{sec-2d-pic}
\begin{figure}[tphb]
	\centering
	\includegraphics{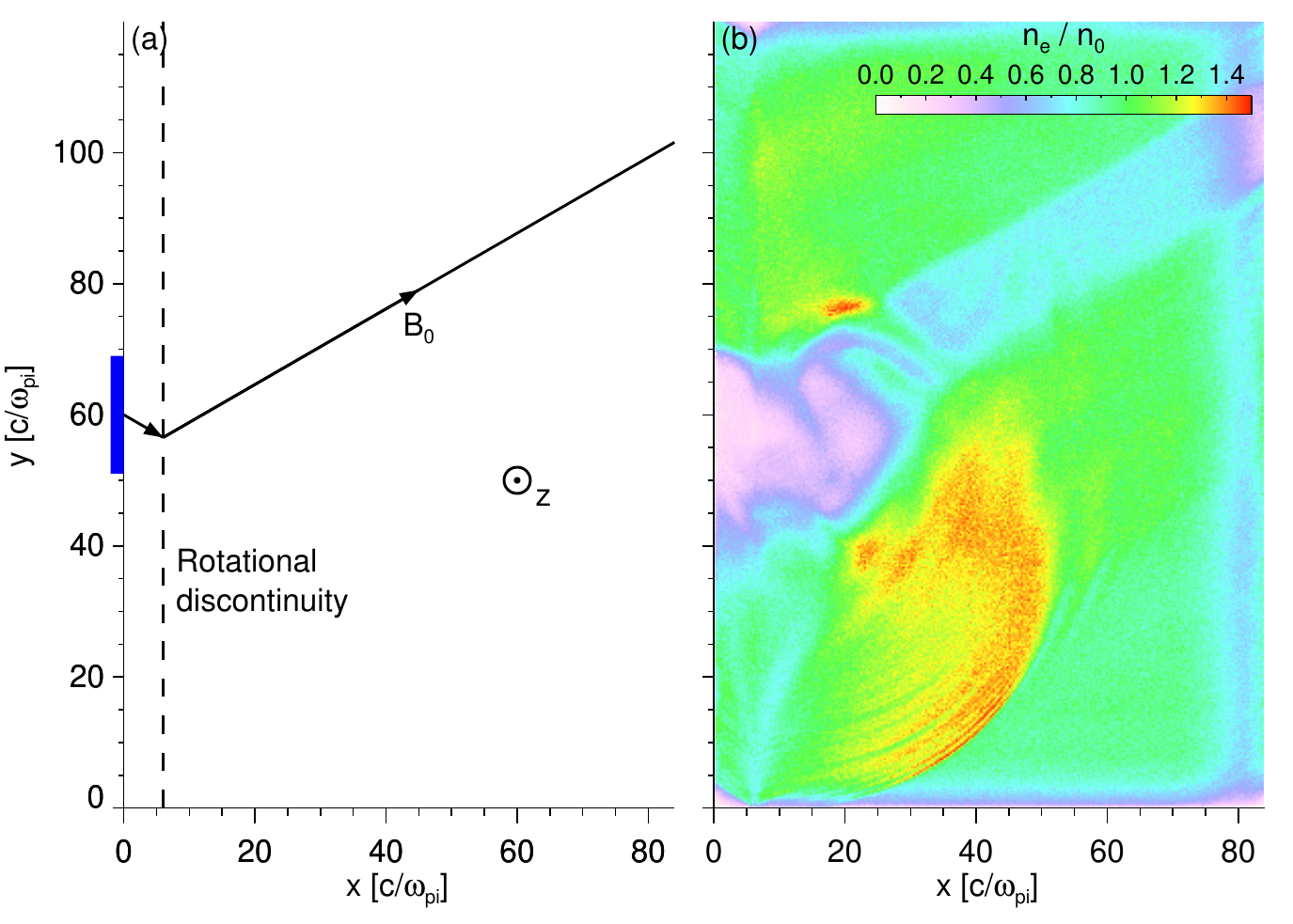}%
	\caption{\label{fig-2d-setup}Formation of foreshock transients in the 2D PIC simulation. (a) The computational setup. (b) The total electron density. The secondary shock, the sheath and the cavity are clearly visible. This and other snapshots hereafter from 2D simulations are taken at the end of the simulation $t = 35.3\, \omega_{ci}^{-1}$.}
\end{figure}

In a more realistic configuration, foreshock ions interact with different types of discontinuities (e.g., tangential discontinuities or rotational discontinuities). The resulting HFAs and FBs expand in both perpendicular and parallel directions. To account for such a scenario, we perform a PIC simulation with two spatial dimensions. The computational setup is shown in Figure \ref{fig-2d-setup}(a). We use perfectly matched layers as absorbing boundary conditions for electromagnetic fields \citep{vay2000new} and also absorbing boundary conditions for particles. The orientation of the background magnetic field $\mathbf{B}_0$ changes from $-30^\circ$ to $+30^\circ$ with respect to the $x$-axis at $x =6\, c/\omega_{pi}$, which gives a rotational discontinuity. The simulation is in the solar wind rest frame, so the discontinuity is not moving. Initially, both ambient ions and electrons have a uniform density $n_0$ in the computation domain. To make computational cost of the 2D simulation affordable, the ion-to-electron mass ratio is $m_i / m_e = 25$, and the normalized Alfv\'en velocity is $v_A / c = 1 / 150$. The initial thermal velocities of ambient ions and electrons are $v_{Ti} = 0.15 v_A$ and $v_{Te} = 1.5 v_A$, respectively. Energetic ions are continuously injected from a cathode on the boundary $x=0$ (see the blue strip located at $51\, c/\omega_{pi} \leqslant y \leqslant 69\, c/\omega_{pi}$ in Figure \ref{fig-2d-setup}(a)). Injected ions stream along $\mathbf{B}_0$ with an initial beam velocity $v_{bi} = 4.5 v_{A}$. The density of injected ions is $0.2 n_0$, and their mass is $m_{bi} = 100 m_e$. To maintain the continuous injection of ions, electrons are injected simultaneously with ions, otherwise the initially injected ions will create a space charge potential that prevents further injection of ions. Injected electrons have a density of $0.2 n_0$ and an initial beam velocity of $1.5 v_A$. Using this setup, a secondary shock and a cavity is developed as shown in Figure \ref{fig-2d-setup}(b), and the structure is qualitatively consistent with that from the recent 3D global hybrid simulations \citep{wang2020evolution}. Below we examine the magnetic field structure and the associated current system of the secondary shock and the cavity, and discuss similarities and differences between the 2D and 1D results.

\begin{figure*}[tphb]
	\centering
	\includegraphics[width=6.75in]{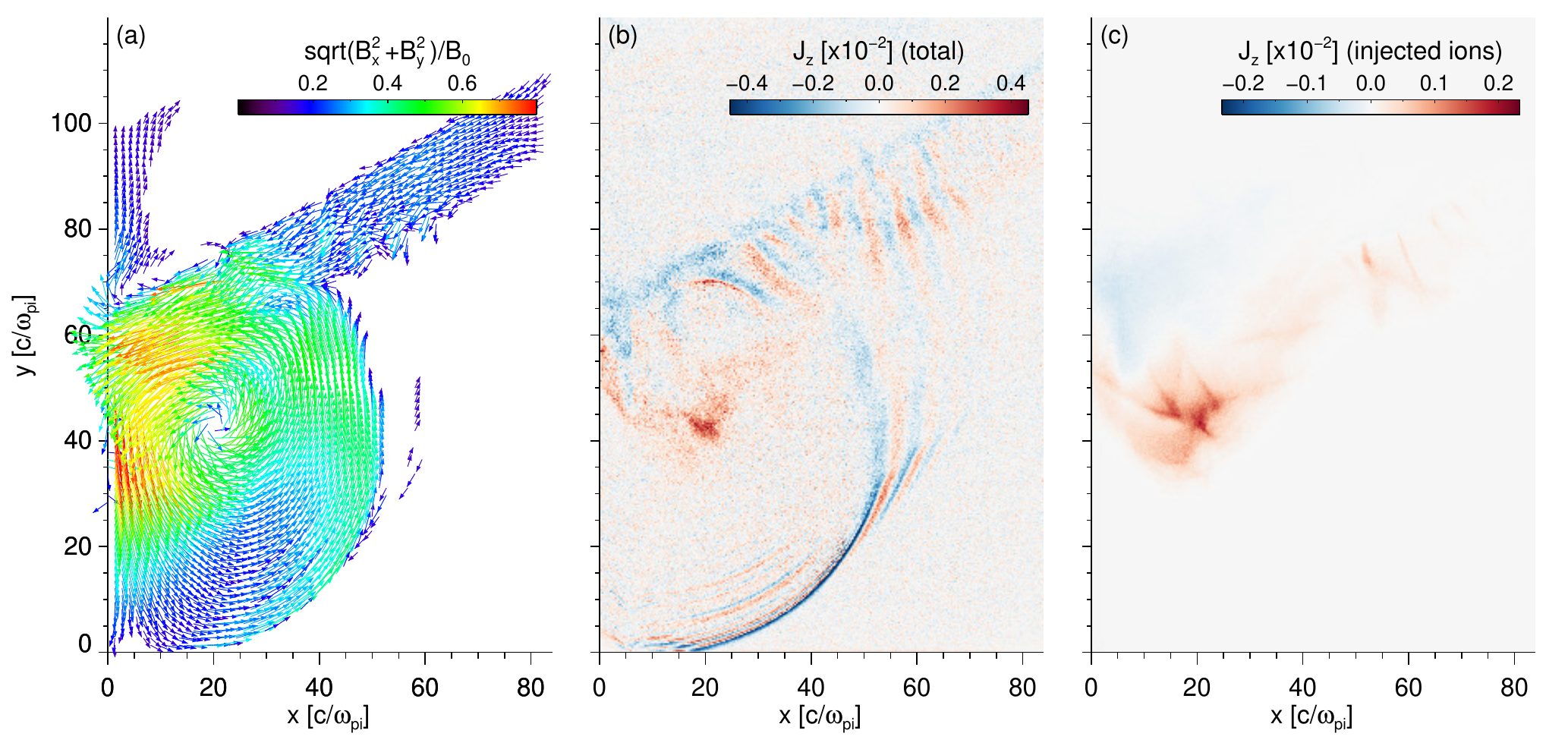}%
	\caption{\label{fig-bxy-jz}In-plane magnetic field and associated currents. (a) The vector field $(B_x, B_y)$. The color scale represents the normalized strength of the in-plane magnetic field $\sqrt{B_x^2 + B_y^2} / B_0$. (b) The total electric current in the $z$ direction. (c) The current $J_z$ contributed by the injected ions.}
\end{figure*}

\begin{figure}[tphb]
	\centering
	\includegraphics[width=6.75in]{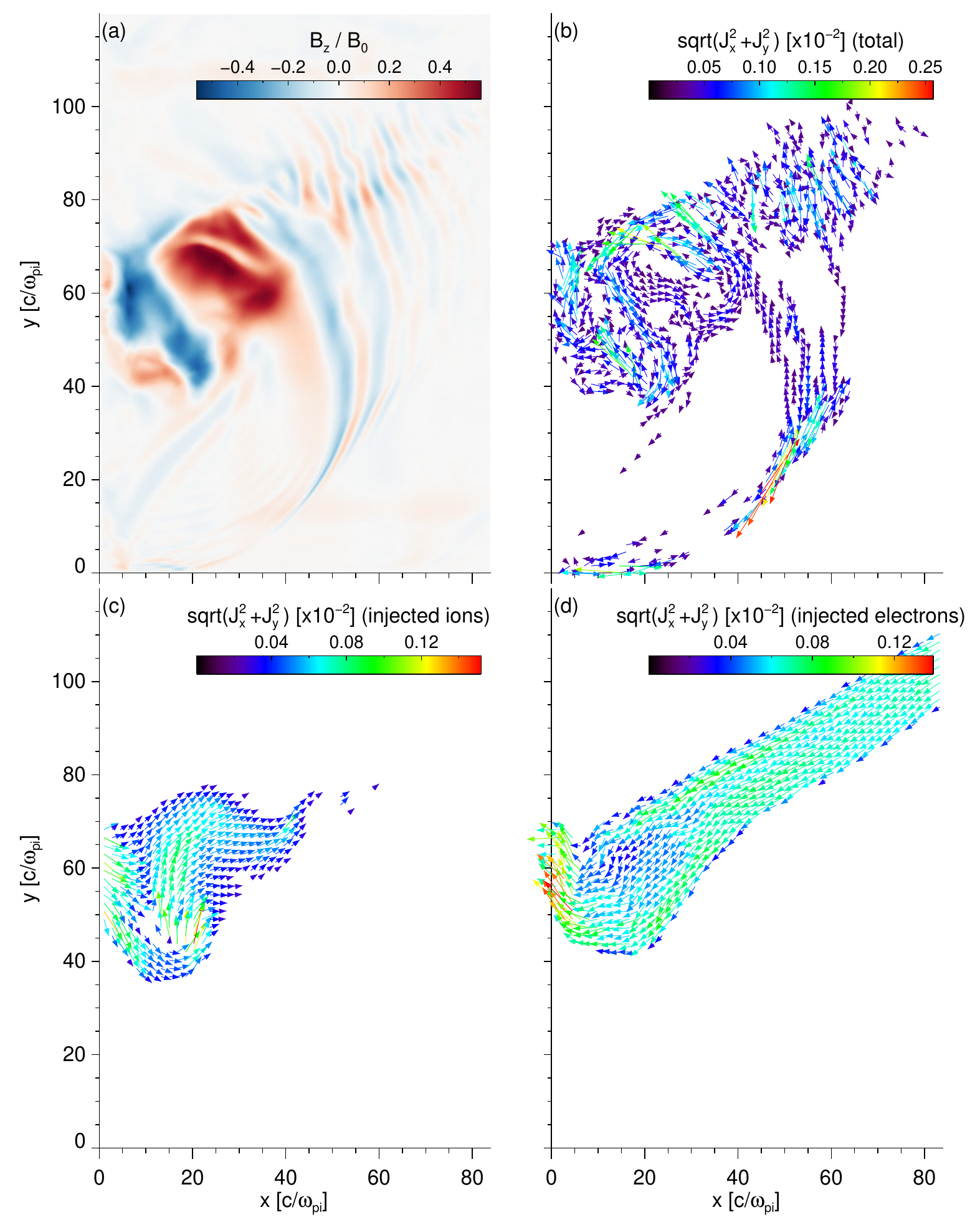}%
	\caption{\label{fig-bz-jxy}Out-of-plane magnetic field and associated currents. (a) The color map of $B_z$. (b) The vector field $(J_x, J_y)$. The color scale represents the normalized strength of the in-plane current $\sqrt{J_x^2 + J_y^2}$. (c), (d) The in-plane currents contributed by injected ions and electrons, respectively.}
\end{figure}

Because the discontinuity spatial scale is far smaller than the gyroradius of injected ions and the timescale of transient formation is comparable to ion gyroperiod, injected ions are demagnetized [Figs.~\ref{fig-bxy-jz}(c) and \ref{fig-bz-jxy}(c)], while injected electrons are nearly always magnetized ($\rho_e \approx 0.052\, c/\omega_{pi}$) and move along the magnetic field lines across the discontinuity [Fig.~\ref{fig-bz-jxy}(d)]. As injected ions cross the rotational discontinuity, they cannot change their velocities immediately so that the injected ion velocity remain in the $x$-$y$ plane near the discontinuity ($x=6$ -- $10\, c/\omega_{pi}$, $y\approx 40\, c/\omega_{pi}$ in Fig.~\ref{fig-bz-jxy}(c)). This velocity projects to both perpendicular and parallel directions, resulting in a pitch angle about $60^\circ$ with respect to $\mathbf{B}_0$. As injected ions move further away from the discontinuity, they gradually gyrate to the $z$ direction ($x=15$ -- $25\, c/\omega_{pi}$, $y\approx 45\, c/\omega_{pi}$ in Fig.~\ref{fig-bxy-jz}(c)). The motions of demagnetized injected ions and magnetized electrons lead to a large-scale Hall current in the $z$ direction [Fig.~\ref{fig-bxy-jz}(b)] and $x$-$y$ plane [Fig.~\ref{fig-bz-jxy}(b)] (also see the contribution from ambient plasma in Appendix \ref{appendix-efields-currents}), which should be maintained by the continuous particle injection. Therefore, a vortex-like vector field of $(B_x, B_y)$ is generated by the $z$ component of the Hall current (the center of the vortex is located where $J_z$ of injected ions is maximized in Figures \ref{fig-bxy-jz}(b) and \ref{fig-bxy-jz}(c)). This gives a region of magnetic cavity on one side [Fig.~\ref{fig-bxy-jz}(a)] and a region of enhanced magnetic field on the other side, which steepens into a secondary shock (see another current at its surface carried by ambient plasma at $x = 0$ -- $60\, c/\omega_{pi}$, $y = 0$ -- $40\, c/\omega_{pi}$ in Figs.~\ref{fig-bxy-jz}(b) and \ref{fig-bz-jxy}(b)). The $x$ and $y$ components of the Hall current [Fig.~\ref{fig-bz-jxy}(b)] generate a bipolar field of $B_z$ [Fig.~\ref{fig-bz-jxy}(a)], which is located inside the cavity and can reach $\pm 0.5 B_0$. There is no such a bipolar $B_z$ in 1D simulations, because the spatial variation of electric current in the $x$-$y$ plane was not resolved. Consequently, the profile of magnetic field strength is not exactly the same as the density profile in 2D, which is more consistent with observations.

The electromagnetic fields caused by the perpendicular (gyrating) and parallel (streaming) motions of injected ions have distinctively different characteristics. On the one hand, driven by the gyrating motion of injected ions, a compressional boundary with enhanced plasma density and magnetic field is formed [Figs.~\ref{fig-2d-setup}(b) and \ref{fig-bxy-jz}(a)] and steepens into a shock (the Mach number $M_f = 1.5$). The formation mechanism is similar to what we have learned from 1D simulations in Section \ref{sec-1d-pic} (also see Appendix \ref{appendix-efields-currents} for more details of electric fields and currents). On the other hand, the streaming motion of injected ions in the parallel direction excites ion beam instabilities [Fig.~\ref{fig-bxy-jz}(a), \ref{fig-bxy-jz}(b), \ref{fig-bz-jxy}(a) and \ref{fig-bz-jxy}(b)]. The waves co-propagate with the ion beam and are right-hand polarized. These waves are likely fast magnetosonic waves excited by injected ion beam through the anomalous cyclotron resonance $\omega - k_\parallel v_\parallel = -\omega_{cb}$ \citep[e.g.,][]{wilson2016low, weidl2019three}, where $v_\parallel \approx v_{bi} \sin 30^\circ = 2.25 v_A$ is the parallel velocity of injected beam ions and $\omega_{cb}$ is the cyclotron frequency of beam ions. Such wave signatures have also been seen in the recent 3D global hybrid simulations by \citet{wang2020evolution}. Because part of the free energy is released in the parallel expansion, the expansion speed of foreshock transients would decrease in comparison with that of 1D simulations.

\section{Conclusions and discussion}\label{sec-conclusions-discussions}
We have shown the detailed process of diamagnetic cavity and secondary shock formation at foreshock transients. Our results provide clear evidence of the critical role electrostatic and induction electric fields play in this formation process, and reveal the energy transfer between different particle species and electromagnetic fields in the formation. Our study also demonstrates how a rotational discontinuity interacts with foreshock ions and leads to the formation of a foreshock bubble. Such a process is not simply increasing the thermal pressure of foreshock ions as explained in the previous models, but demagnetizing foreshock ions and generating the Hall current. Our ensemble of simulations indicates that the expansion speed of foreshock transients is proportional to the hot ion density ratio and thermal speed, suggesting that foreshock transients with secondary shocks are more prevalent at high Mach number astrophysical shocks than those already observed at planetary bow shocks. Since these ion-kinetic structures have been shown to accelerate particles cooperatively with primary shocks with high efficiency \citep{wilson2016relativistic, liu2017fermi, liu2018ion, turner2018autogenous, liu2019relativistic}, they could significantly contribute to high Mach number astrophysical shock acceleration, e.g., the generation of the cosmic rays. Therefore, they must be included in shock acceleration models in general.

In our 2D simulation, we only consider a rotational discontinuity with a certain magnetic field configuration. The basic physical process, however, is general. Different types of discontinuity and magnetic field configurations affect the details of, e.g., how foreshock ions are demagnetized and how the corresponding Hall current changes the background magnetic field. In the future, these details will be examined by more advanced simulations. By collaborating with the results from our parameter scan, a model might be established to predict the formation of foreshock transients and secondary shocks given certain upstream conditions and discontinuity configurations.

\acknowledgments

We are grateful to J.~Hohl for editing an early version of the manuscript. We would like to acknowledge high-performance computing support from Cheyenne (doi:10.5065/D6RX99HX) provided by NCAR's Computational and Information Systems Laboratory, sponsored by the National Science Foundation. We would also like to acknowledge the OSIRIS Consortium, consisting of UCLA and IST (Lisbon, Portugal) for the use of \textsc{osiris} and for providing access to the \textsc{osiris} 4.0 framework. Data access and processing in Figure \ref{setup}(b) was done using SPEDAS V3.1, see \citet{angelopoulos2019space}. XA and JB are supported by the NASA/HTIDS grant NNX16AG21G. TZL is supported by the NASA Living With a Star Jack Eddy Postdoctoral Fellowship Program, administered by the Cooperative Programs for the Advancement of Earth System Science (CPAESS). TZL and XA are partially supported by NSF award AGS-1941012. The supplemental videos can be accessed via Zenodo \url{https://doi.org/10.5281/zenodo.3951168}.

\appendix
In this appendix, we provide supporting information for the main manuscript. In the first section, we describe the details of the simulation setup. In the second section, we identify the contribution of hot and ambient ions to the diamagnetic current, which complements the contribution of the electrons shown in the main manuscript. In the third section, we provide details of the parameter scan and discuss the saturation time of foreshock transients. In the fourth section, we use spacecraft observations to support the conclusion that a large Mach number of the primary shock favors foreshock transient formation. In the fifth section, we present details of electric fields and currents for 2D PIC simulations.

\section{Computational setup}\label{appendix-computational-setup}
Below we present the details of the computational configuration, how the parameters are scaled from the space environment to the numerical experiments, and the effects of including more spatial dimensions.

\subsection{Configuration}
\begin{figure}[tphb]
 \centering
 \includegraphics[width=4in]{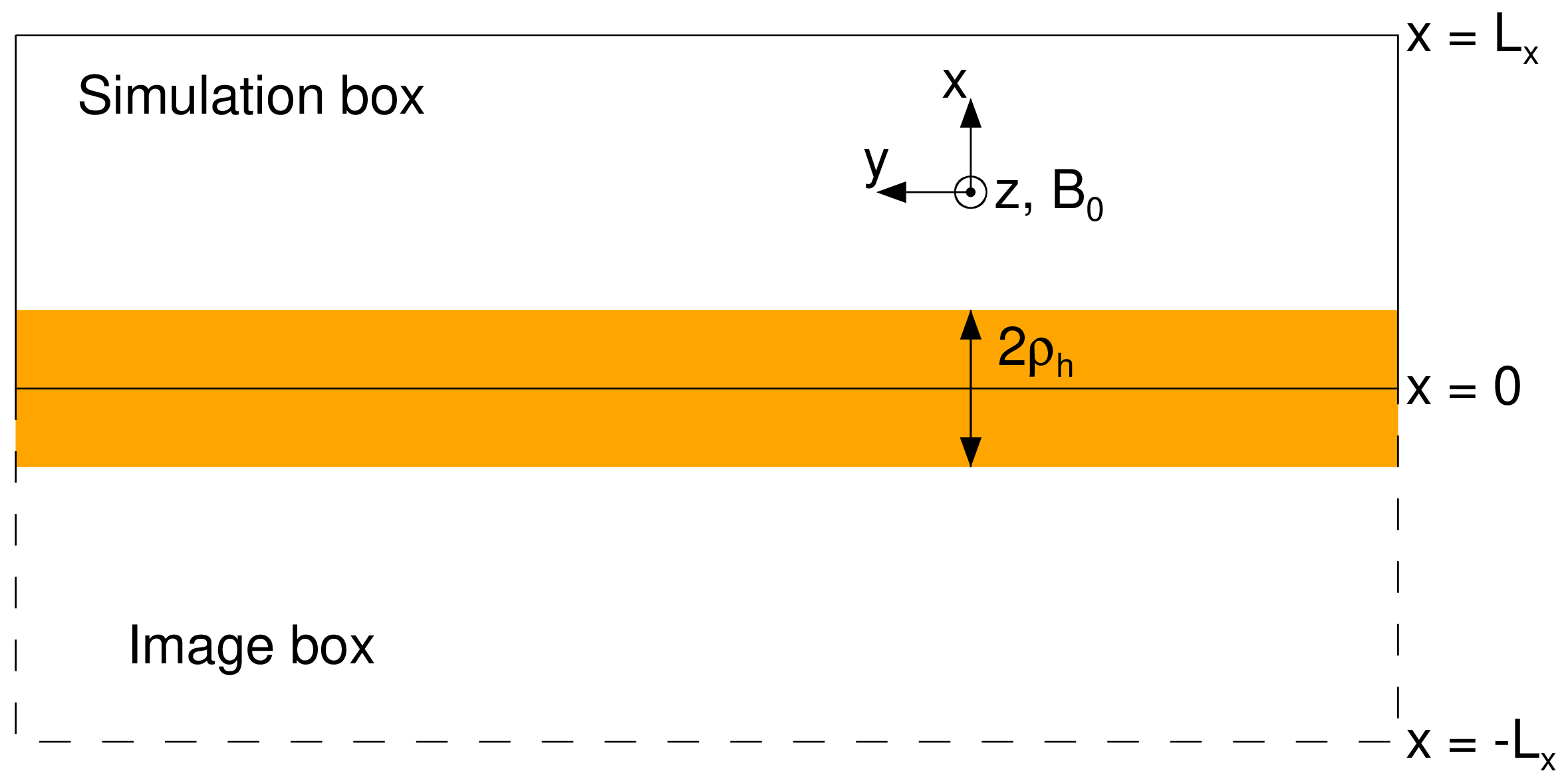}%
 \caption{\label{simulation-setup}A sketch of the simulation box. See the text for details of the simulation setup.}
 \end{figure}

We used the massively parallel, fully electromagnetic PIC code \textsc{osiris} \citep{fonseca2002osiris, hemker2015particle} for our simulations. The simulations have one dimension ($x$) in configuration space and three dimensions ($v_x, v_y, v_z$) in velocity space. The computational domain is $-L_x \leqslant x \leqslant L_x$ [Fig.~\ref{simulation-setup}]. The specific size of the system $L_x$ is chosen based on the condition of the hot ions as described below. The cell length $\Delta_x$ is $2 \lambda_D$, where $\lambda_D = v_{Te} / \omega_{pe}$ is the initial electron Debye Length, $v_{Te}$ is the electron thermal velocity, and $\omega_{pe}$ is the electron plasma frequency. Each cell contains $500$ particles per species. The time step is set as $0.95 \Delta_x / c$ to satisfy the Courant–Friedrichs–Lewy condition in one dimension, where $c$ is the speed of light. The ambient magnetic field $B_0$ is along the $+z$ direction. The electron cyclotron frequency $\omega_{ce}$ is equal to $\omega_{pe} / 30$. Given the reduced ion-to-electron mass ratio $m_i / m_e = 100$, the Alfv\'en velocity is $v_A = c / 300$.

To mimic the concentrated hot ion core in foreshock transients, a mixture of hot and ambient ions is initialized in the layer $-\rho_h \leqslant x \leqslant \rho_h$ [Fig.~\ref{simulation-setup}], where $\rho_{h}$ is the gyroradius corresponding to the initial thermal velocity $v_{Th}$ of hot ions. Inside the hot layer, the densities of hot and ambient ions are $\eta n_0$ and $(1 - \eta) n_0$, respectively, where $\eta$ denotes the fraction of hot ions. Outside the hot layer, the only positively charged species is ambient ions with density $n_0$. Electrons are initialized with a uniform density $n_0$ in the entire domain. The initial thermal velocities of ambient electrons and ions are $v_{Te} = 3 v_A$ and $v_{Ti} = 0.3 v_A$, respectively. In the nominal run presented in the main manuscript, the hot ion concentration ratio is $\eta = 0.2$, and the initial thermal velocity of hot ions is $v_{Th} = 9 v_A$. To explore the conditions under which secondary shocks are formed, we complete $25$ runs (including the nominal run), varying the hot ion thermal velocity $v_{Th}$ and the hot ion concentration ratio $\eta$. In each run, the hot ion thermal velocity is varied as $v_{Th} = (12v_A, 9v_A, 6v_A, 3v_A, v_A)$ for a given hot ion concentration ratio in the sequence $\eta = (0.05, 0.1, 0.2, 0.5, 1.0)$; other parameters are kept unchanged. The distance over which the secondary shock saturates critically depends on the typical hot ion gyroradius $\rho_h$. For each hot ion thermal velocity in the sequence $ v_{Th} = (12v_A, 9v_A, 6v_A, 3v_A, v_A)$, the corresponding domain size is chosen as $L_x = (36d_i, 72d_i, 108d_i, 144d_i, 180d_i)$, such that the full evolution of the secondary shocks can be resolved within the computation domain. Here $d_i$ denotes the ion inertial length.

The domain's upper half $0 \leqslant x \leqslant L_x$ (simulation box) and lower half $-L_x \leqslant x \leqslant 0$ (image box) are symmetrical to its center $x = 0$. Although the additional image box doubles the computational cost, it allows us to use the periodic boundary condition for both fields and particles. When presenting the results, we focus on the simulation box. 

\subsection{Scale-down numerical experiments}
Table \ref{scale-down} shows how the parameters are scaled from Earth's foreshock to the nominal simulation. The absolute values of the thermal velocities of different species and the alfven velocity in the numerical experiments are larger than those in the space environment, for the sake of saving computing time. But the key dimensionless parameters are the scale separations between the gyro-radii of electrons, ambient ions and hot ions, because they determine the charge separation and hence the electrostatic fields. We keep the ratio between the gyro-radii of different species on the same order as the measured values to capture the key ingredient for foreshock transient formation.

\begin{table}[tphb]
\centering
\begin{tabular}{l *{4}{c}}
 \hline
      [\text{km/s}]          & $v_A$   & $v_{Te}$ & $v_{Ti} $ & $v_{Th}$ \\
 \hline
Earth's foreshock   & $50$  & $2000$ & $50$ & $400-700$ \\
nominal simulation   & $1000$    & $3000$  & $300$ & $9000$ \\
 \hline
               & $v_A / c$ & $m_i / m_e$ & $\rho_i / \rho_e$ & $\rho_{h} / \rho_e$ \\
\hline
Earth's foreshock   & $1/6000$  & $1836$ & $45$ & $360 - 630$ \\
nominal simulation   & $1/300$    & $100$  & $10$ & $300$ \\
 \hline
\end{tabular}
\caption{\label{scale-down}The comparison of relevant dimensional and non-dimensional parameters between Earth's foreshock and the nominal simulation. }
\end{table}

\subsection{Role of Alfv\'en speed}
To make the computational cost affordable, we have chosen a larger Alfv\'en speed $v_A$ than the realistic value in the simulations. Thus it is necessary to justify how $v_A$ affects the formation process of foreshock transients. The electrostatic field is caused by a small charge separation. This small charge separation is induced by the difference between the gyro-radii of hot ions and other species. To estimate the magnitude of the electrostatic field, we invoke the Gauss's Law
\begin{eqnarray}
\frac{E}{\rho_h} \sim 4 \pi \epsilon e n_h , \nonumber
\end{eqnarray}
where $\rho_h$ can be further replaced with $v_{\text{Th}} / \omega_{ci}$, and $\omega_{ci}$ is the ion cyclotron velocity. $\epsilon$ is a small dimensionless parameter standing for the ``small'' charge separation ($\epsilon \ll 1$). Let us make the notation
\begin{eqnarray}
n_h = \alpha \cdot n_0 , \nonumber \\
v_{\text{Th}} = \beta \cdot v_A . \nonumber
\end{eqnarray}
The electrostatic field can be rewritten as 
\begin{eqnarray}
E \sim \epsilon \alpha \beta \cdot \frac{4 \pi e n_0 v_A}{\omega_{ci}} = \epsilon \alpha \beta \cdot \frac{\omega_{pi}^2}{\omega_{ci}^2} \frac{v_A}{c} B = \epsilon \alpha \beta \cdot \frac{c}{v_A} B . \nonumber
\end{eqnarray}
Here $\omega_{ci} = B e / m_i c$ and $\omega_{pi}^2 = 4 \pi n_0 e^2 / m_i$. We have also made use of $v_A / c = \omega_{ci} / \omega_{pi}$. It is more appropriate to normalize the electrostatic field as
\begin{eqnarray}
\left( \frac{c E}{B} \right) / v_A \sim \epsilon \alpha \beta \left( \frac{c}{v_A} \right)^2 , \nonumber 
\end{eqnarray}
which is also the ratio of electric drift velocity to Alfv\'en velocity. As $v_A$ decreases, the electrostatic field in fact increases with the scaling $(c / v_A)^2$. Intuitively, as the strength of magnetic field decreases (lower $v_A$), the difference between the gyro-radii of hot ions and electrons becomes larger, inducing stronger charge separation and larger electrostatic field. Indeed, in \citet{liu2017statistical}, it is seen that lower interplanetrary magnetic (IMF) field has higher probability to form foreshock transients. Therefore lower $v_A$ favors the formation of foreshock transients. In this sense, the secondary shocks in reality may form more easily than in our simulations.


\section{Diamagnetic current}\label{appendix-diamagnetic-current}
\begin{figure}[tphb]
	\centering
	\includegraphics{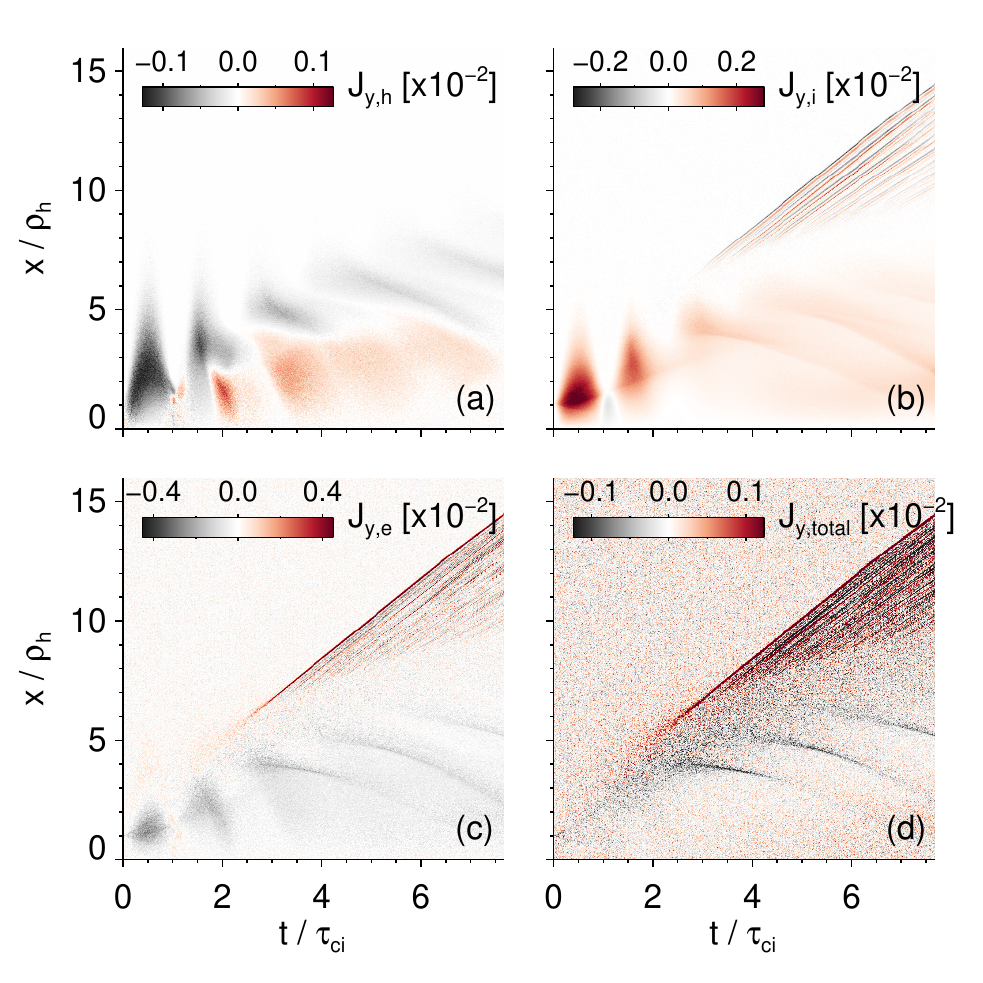}%
	\caption{\label{ion-current}The spatio-temporal evolution of total current contributed by different species. (a) Electric current in the $y$ direction contributed by hot ions. (b)  Electric current in the $y$ direction contributed by ambient ions. (c) Electric current in the $y$ direction contributed by electrons. (d) Total electric current in the $y$ direction contributed by all three species. To emphasize the current patterns, the color scales differ in each panel.}
\end{figure}

To show the contribution of each species to the total current that shapes the magnetic cavity and compression in foreshock transients, we plot the current $J_y$ by each species and the total current in Fig.~\ref{ion-current}. In the first and second gyrations of the hot ions ($0 < t /\tau_{ci} \lesssim 2$), a hot ion current in the $-y$ direction consistent with the sense of ion gyration is seen [Fig.~\ref{ion-current}(a)]. The ambient ions, on the other hand, form a current in the $+y$ direction [Fig.~\ref{ion-current}(b)] because more ambient ions are distributed outside the hot layer than inside it. As explained in the main manuscript, electrostatic fields $E_x$ cause an electron current in the $-y$ direction to form [Fig.~\ref{ion-current}(c)]. The electron current and the hot ion (partial diamagnetic) current in the $-y$ direction dominate the ambient ion (partial diamagnetic) current in the $+y$ direction. This results in a net Hall current [Fig.~\ref{ion-current}(d)] that reduces the magnetic field on one side (the cavity) and enhances it on the other side (the compressional boundary). As the magnetic cavity is developed and continues to expand ($ \tau_{ci} \gtrsim 2$), hot ions can gyrate out farther [Fig.~\ref{ion-current}(a)]. As a result, the hot ion current is along the $-y$ direction at the leading edge of the cavity as some hot ions gyrate outward and is along the $+y$ direction inside the cavity as some other hot ions gyrate inward. The gyroradii of ambient ions and electrons are much smaller than the characteristic spatial scale of the electrostatic fields. This gives rise to similar electric current profiles for ambient ions and electrons [Fig.~\ref{ion-current}(b) and \ref{ion-current}(c)] caused by the drift $-c E_x / B_z$. The current associated with the magnetic perturbations of the magnetosonic waves is also seen [Fig.~\ref{ion-current}(b), \ref{ion-current}(c) and \ref{ion-current}(d)].

\section{Parameter scan}\label{appendix-parameter-scan}
\begin{figure}[tphb]
	\centering
	\includegraphics[width=6.75in]{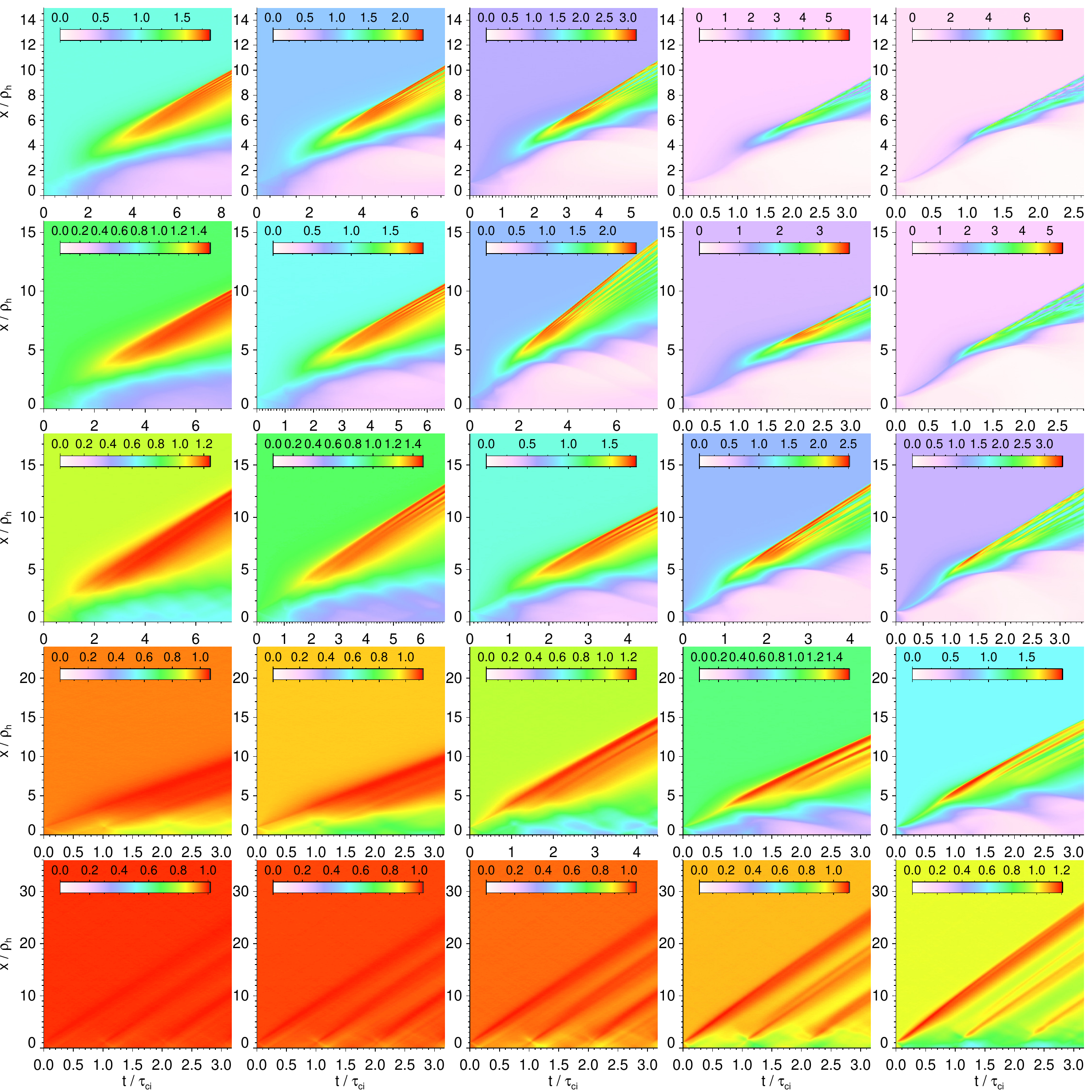}%
	\caption{\label{mag-mosiac2}The spatio-temporal evolution of the normalized magnetic field $B_z / B_0$ for the parameter scan. From top to bottom, each row corresponds to the initial thermal velocity of hot ions in the sequence $v_{Th} = (12v_A, 9v_A, 6v_A, 3v_A, v_A)$. From left to right, each column corresponds to the hot ion concentration ratio in the sequence $\eta = (0.05, 0.1, 0.2, 0.5, 1.0)$. The specific value of the normalization unit $\rho_h$ is different for each row. The total simulation time is different for each panel, which is chosen to resolve the full evolution of the system.}
\end{figure}

\begin{figure}[tphb]
	\centering
	\includegraphics[width=6.75in]{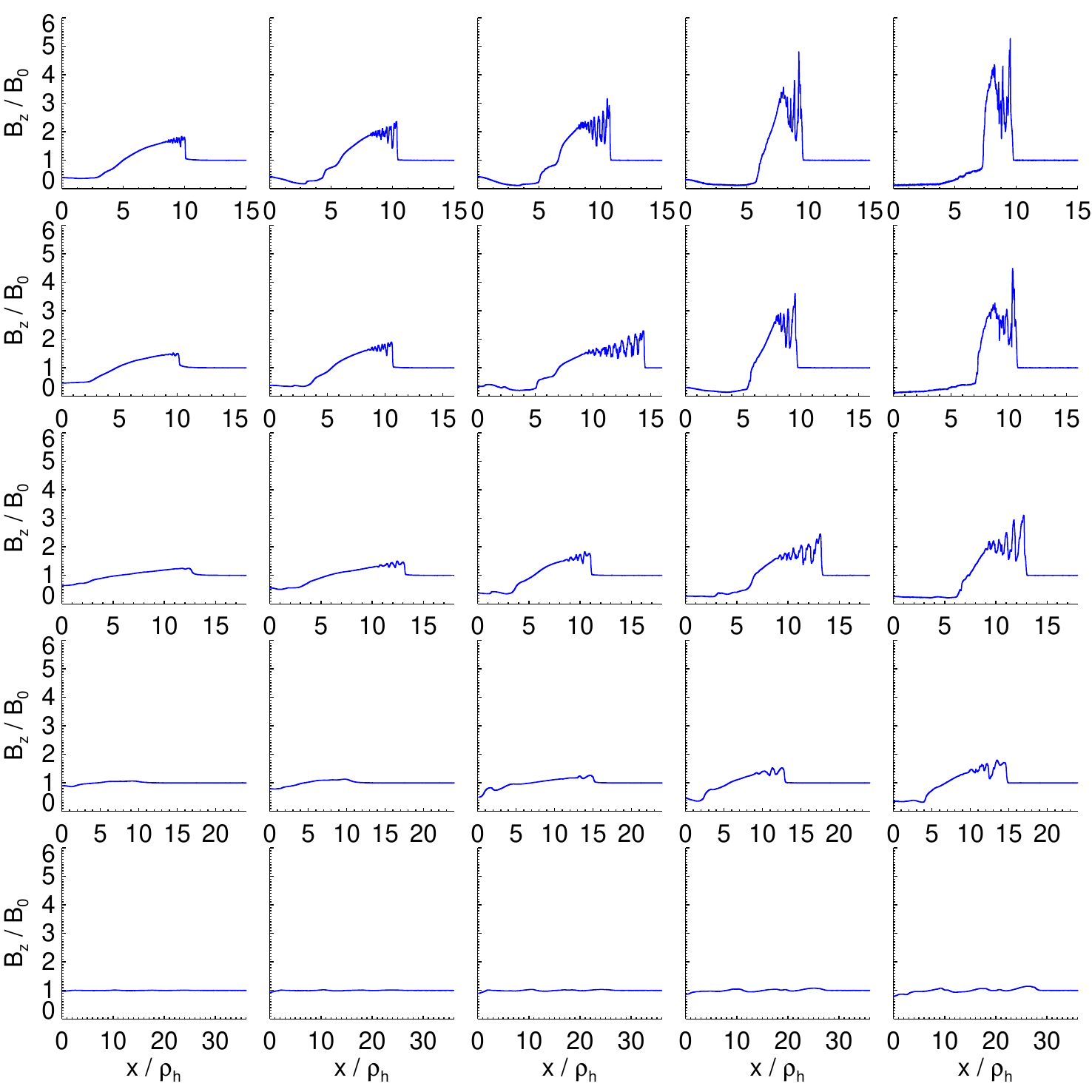}%
	\caption{\label{mag-mosiac}The spatial profiles of the magnetic field $B_z / B_0$ for the parameter scan. The position of each panel is arranged in the same way as in Fig.~\ref{mag-mosiac2}. Each profile is taken at the end of the simulation.}
\end{figure}

In Fig.~\ref{mag-mosiac2}, we show the detailed evolution of the magnetic field for $25$ runs in the parameter scan. Correspondingly, Fig.~\ref{mag-mosiac} shows the spatial profile of the magnetic field at the end of each run. As the hot ion thermal velocity ($v_{Th}$) and the hot ion concentration ratio ($\eta$) increase, depletion of the magnetic flux in the foreshock cavity, as well as pile up of the magnetic flux in the compressional boundary, becomes more enhanced. At the same time, the Mach number of secondary shocks increases with $v_{Th}$ and $\eta$. For a higher Mach number of the primary shock, higher thermal velocity and higher concentration ratio of foreshock ions/hot ions will be produced \citep{burgess2012ion, paschmann1983ion}. Thus, the parameter scan strongly indicates that foreshock transients and associated secondary shocks are more likely formed at the high Mach number shocks.

The saturation time $t_s$ of the foreshock transients, i.e., the time when energy transfer between particles and fields vanishes, can be estimated as $\rho_{\text{eff}} / v_{\text{shock}}$. Here $v_{\text{shock}} = M_f v_f$ is the propagation speed of the secondary shock, and $\rho_{\text{eff}}$ is the effective gyroradius of hot ions. The parameter $\rho_{\text{eff}}$ characterizes the typical length over which the compressional boundary is detached from the hot ions and takes the value $C \rho_h$, where $C$ is a constant that depends primarily on the hot ion distribution. For the Maxwellian distribution used in this study, $C$ takes the approximate value $5$ based on all the runs we perform. Therefore, we have $$t_s = \frac{C}{\omega_{ci}} \cdot \frac{v_{Th}}{M_f (v_{Th}, \eta) \cdot v_f} ,$$ where we have written $\rho_h$ as $v_{Th} / \omega_{ci}$. The fast magnetosonic Mach number $M_f$ is a function of both $v_{Th}$ and $\eta$. For a given hot ion thermal velocity, the saturation time ($t_s \propto 1 / M_f$) decreases as the hot ion concentration ratio increases. For a given hot ion concentration ratio, the saturation time ($t_s \propto v_{Th} / M_f$) slowly increases as the hot ion thermal velocity increases, indicating the dependence $M_f \propto v_{Th}^{\gamma}$ ($\gamma < 1$).

Of interest is that, in the last row of Fig.~\ref{mag-mosiac2} with $v_{Th} = v_A$, hot ions generate a small magnetic perturbation each time when they gyrate out of the hot layer. These magnetic perturbations propagate at the fast magnetosonic speed $v_f$. In these cases, hot ions do not have enough time to transfer their free energy to the magnetic field in each interaction, the remaining free energy leads to the consecutive generation of propagating magnetic perturbations.

\section{Correlation between the occurrence rate of foreshock transients and primary shock's Mach number}\label{appendix-correlation}
\begin{figure}[tphb]
 \centering
 \includegraphics[width=3.375in]{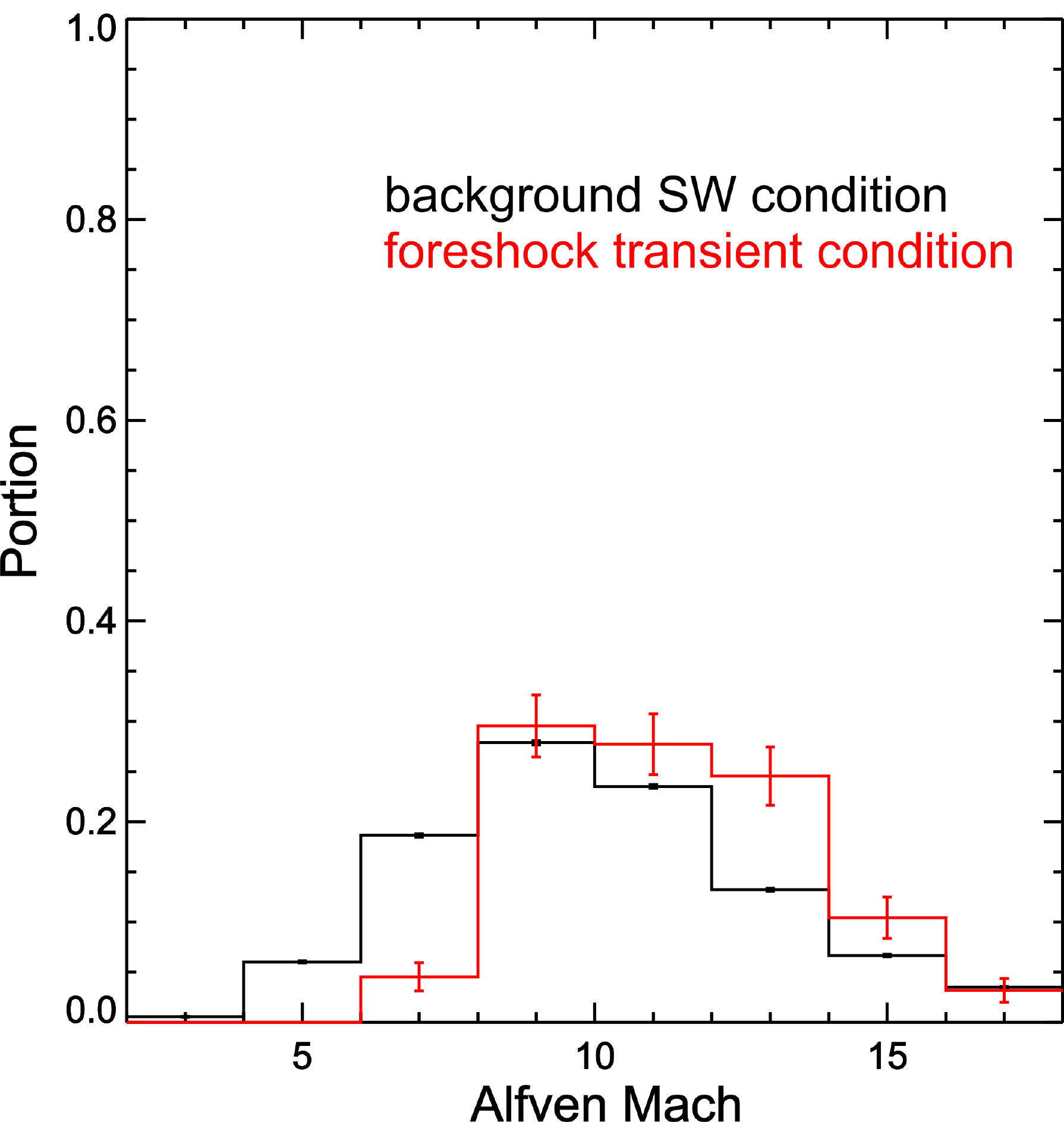}%
 \caption{\label{ma-stat}The probability distribution of Earth's bow shock Alfven Mach number. The black bars and red bars represent the probability distribution during the entire time interval of the database and during each foreshock transient observation, respectively.}
 \end{figure}

The statistical study by \citet{liu2017statistical} shows that high solar wind speed and low interplanetary magnetic field strength favor the formation of foreshock transients at Earth's bow shock, implying that a large solar wind Alfv\'en Mach number is a favorable condition. To confirm this, we use the same database as in \citet{liu2017statistical} and plot the probability distribution of the solar wind Alfv\'en Mach number during the entire time interval of the database (black) and during each observation of a foreshock transient (red), as shown in Fig.~\ref{ma-stat}. We see that the latter one indeed shows higher probability at large Alfv\'en Mach number than the former one. This result indicates that a large primary shock Mach number favors the formation of foreshock transients. Further statistical study is needed to test the relationship between the primary shock Mach number and secondary shock formation. 

\section{Electric currents by ambient particles and Electric fields in 2D PIC simulations}\label{appendix-efields-currents}

\begin{figure}[tphb]
	\centering
	\includegraphics[width=6.75in]{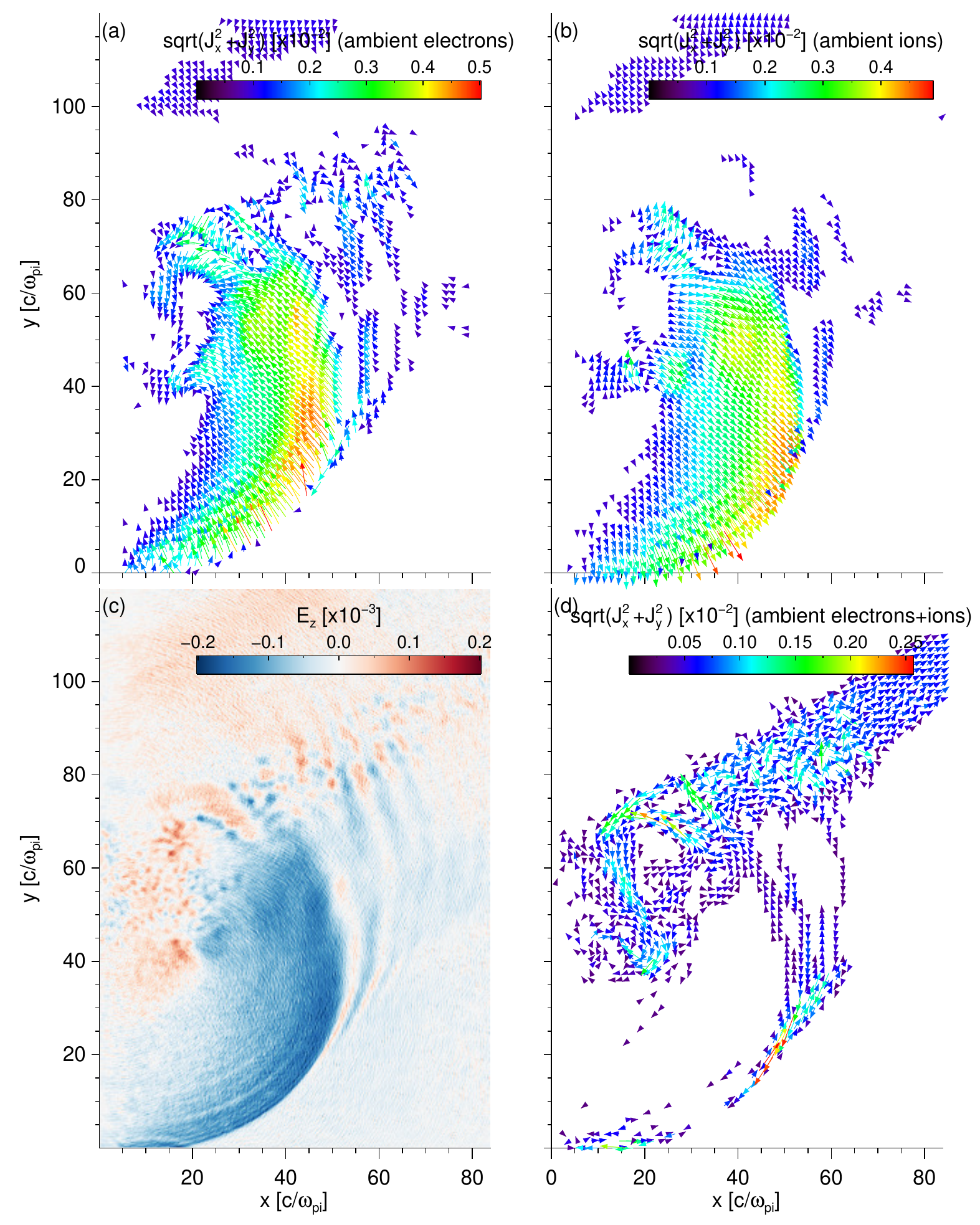}%
	\caption{\label{fig-jxy-ez}(a), (b) The vector fields of $(J_x, J_y)$ contributed by ambient electrons and ions, respectively. The color scale represents the magnitude of $\sqrt{J_x^2 + J_y^2}$. (c) The electric field in the $z$ direction. (d) The sum of vector fields of $(J_x, J_y)$ from ambient electrons and ions.}
\end{figure}

The electric field $E_z$ is shown in Figure \ref{fig-jxy-ez}(c), which is an induction electric field coupled with the time-varying $(B_x, B_y)$ of the compressional boundary. This induction electric field has also been observed in 1D simulations in Section \ref{sec-1d-pic}. Ambient elecctrons and ions are advected by this electric field to co-propagate with the shock, as shown in Figures \ref{fig-jxy-ez}(a) and \ref{fig-jxy-ez}(b).

\begin{figure}[tphb]
	\centering
	\includegraphics[width=6.75in]{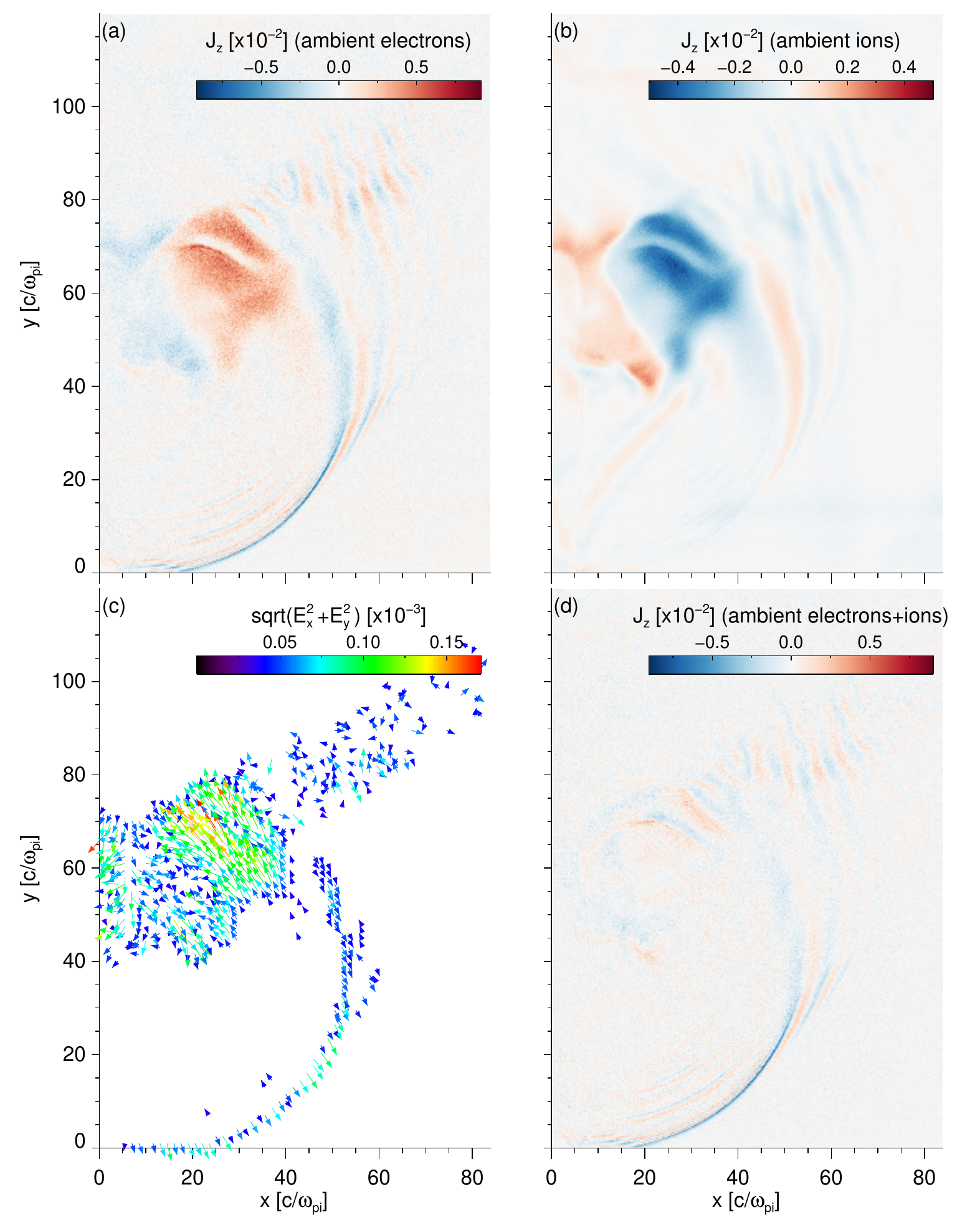}%
	\caption{\label{fig-jz-exy}(a), (b) The currents in the $z$ direction contributed by ambient electrons and ions, respectively. (c) The vector field of $(E_x, E_y)$. The color scale represents the magnitude of $\sqrt{E_x^2 + E_y^2}$. (d) The sum of $J_z$ from ambient electrons and ions.}
\end{figure}

The in-plane electric field $(E_x, E_y)$ is shown Figure \ref{fig-jz-exy}(c). It has the most significant magnitude inside the cavity where strong cross-field current of injected ions is located [Figs.~\ref{fig-bxy-jz}(c) and \ref{fig-bz-jxy}(c)]. Appreciable electric field magnitude $\sqrt{E_x^2 + E_y^2}$ is also distributed in other regions, such as the shock surface and the region of fast magnetosonic waves. Due to the $\mathbf{E} \times \mathbf{B}$ drift caused by the in-plane electric field, ambient electrons and ions have currents $J_z$ similar to 1D simulations, as shown in Figures \ref{fig-jz-exy}(a), \ref{fig-jz-exy}(b) and \ref{fig-jz-exy}(d). However, because the in-plane electric field has both electrostatic and electromagnetic components, and they cannot be easily separated using the present simulation setup, the contribution of electron current due to the electrostatic field is difficult to diagnose. On the shock surface, on the other hand, it is clear to see the static electric field, which gives the surface current on the secondary shock as shown in Figures \ref{fig-jz-exy}(d) and \ref{fig-jxy-ez}(d).

\bibliography{shock}{}
\bibliographystyle{aasjournal}



\end{document}